\documentclass[varenna]{cimento}
\usepackage{graphicx}
\def\PRL{ Phys. Rev. Lett. }

\def\PRB{{Phys. Rev.} B }

\def\be{\begin{equation}}
\def\ee{\end{equation}}
\def\bea{\begin{eqnarray}}
\def\eea{\end{eqnarray}}
\def\ni{\noindent}

\title{COMPOSITE FERMIONS IN QUANTUM HALL SYSTEMS}
\author{
   John J. Quinn$^1$, Arkadiusz W\'ojs$^{1,2}$, 
   Kyung-Soo Yi$^{1,3}$, and Jennifer J. Quinn$^4$}
\institute{$^1$University of Tennessee, Knoxville, TN 37996, U. S. A.\\
   $^2$Wroclaw University of Technology, 50-370 Wroclaw, Poland\\
   $^3$Pusan National University, Pusan 609-735, Korea\\
   $^4$Occidental College, Los Angeles, CA 90041, U. S. A.}
\begin{document}
\maketitle
\vspace{-3cm}
\begin{abstract}
The occurrence of incompressible quantum fluid states of a two dimensional system is a result of 
electron--electron interactions in a highly degenerate fractionally filled Landau level.
Novel quasiparticles (QP's) called composite Fermions (CF's) have allowed a simple single particle
description of the most prominent incompressible states.
Residual interactions among these QP's are investigated.
These interactions determine the type of ``daughter states'' that can occur at the next generation.
We demonstrate that at certain values of the QP filling factor $\nu_{\rm QP}$, Laughlin correlations among
the QP's give rise to states of the standard CF hierarchy.  At other values of $\nu_{\rm QP}$ pairing of QP's
is found leading to a novel hierarchy of incompressible states.
\end{abstract}
%\pacs{71.10.Pm, 73.43.-f}
\vspace{-1cm}
\section*{OUTLINE OF LECTURES}

1. Introduction

2. Electrons Confined to a Two Dimensional Surface in a Perpendicular Magnetic Field

3. Integral Quantum Hall Effect

4. Fractional Quantum Hall Effect

5. Numerical Studies

6. Chern--Simons Gauge Field

7. Jain's Composite Fermion Picture

8. Pseudopotentials

9. Angular Momentum

10. Coefficients of Fractional Parentage

11. Non-Harmonic Pseudopotentials and Correlations

12. Correlations in Higher Landau Levels

13. Chern--Simons Gauge Field Revisited

14. Gedanken Experiments : Laughlin States and the Jain Sequence

15. The Composite Fermion Hierarchy

16. Quasiparticle--Quasiparticle Interactions

17. Quasiparticle--Quasiparticle Pairing and Novel Families of Incompressible States

References

\section{Introduction}
In these three lectures we will try to give a somewhat deeper understanding of concepts commonly
used to interpret experimental data on the quantum Hall effect.
Some of the ideas presented here are neither commonly used nor commonly appreciated by active researches and by referees
for major journals.
In particular, we hope to make clear why the Chern--Simons (CS) mean field (MF) approximation correctly
predicts the structure of the lowest band of energy states for any value of the applied magnetic field $B$, despite introducing
an energy scale that is large but totally irrelevant to the determination of that structure.
We demonstrate that by adding the CS flux to each electron adiabatically (the way Laughlin originally created
Laughlin quasiparticles (QP's) using his imaginary infinitesimal solenoid) instead of via a gauge transformation,
no change in particle statistics results for any value of the added flux, but Laughlin correlations among the
resulting composite Fermions (CF's) arise automatically. 
(CF's are electrons together with their attached CS flux tubes.) 
The equivalence of CF's and Laughlin correlations is important to realize.

The pseudopotential $V(L')$, i.e. the energy of interaction of a pair of electrons, 
each with angular momentum $l$ on a Haldane spherical surface, depends on
the total pair angular momentum $L' = 2l - \mathcal{R}$, where $\mathcal{R}$, called the {\sl relative}
angular momentum of the pair, must be an odd integer.
A very useful by largely unnoticed theorem on pair angular momentum can be used to show that a harmonic pseudopotential of the form
$V_H (L') = A + BL'(L'+1)$, where $A$ and $B$ are constants, introduces no correlations.
For a harmonic potential the energy $E_\alpha (L)$ of every total angular momentum multiplet
$|l^N ; L \alpha >$ is the same for every multiplet that has the same value of $L$.
This means any linear combination of multiplets with the same value of $L$ has the same energy.
Laughlin correlations occur only for {\sl superharmonic} pseudopotentials in which 
$\Delta V(L') = V(L') - V_H(L')$ is positive,
so that large values of $L'$ (and large repulsion) are avoided.
The pseudopotential for higher Landau levels (LL's) is not superharmonic at the largest value of $L'$ (or at the smallest value of
$\mathcal{R}$, $\mathcal{R}=1$). 
Because of this fact Laughlin correlations do not occur for filling factors satisfying $8/3 \geq \nu \geq 7/3$.
Instead, electrons tend to form pairs with the smallest allowed size in order to avoid states with large pair
amplitude at $\mathcal{R}=3$.

The composite Fermion hierarchy of condensed states was based on the reapplication of the CS flux attachment to
QP's in partially filled CF shells.
It predicted (as did the Haldane hierarchy scheme) condensed states at all fractions with odd denominators.
However, it was based on the CF picture being applicable at every level of the hierarchy.
The residual interactions between QP's, $V_{\rm QP-QP}$, have been obtained from numerical studies.
For small values of $\mathcal{R}$, $V_{\rm QP-QP}(\mathcal{R})$ is known reasonably up to an overall constant
(which has no effect on the nature of the correlations).
The nature of the ground state is determined by short range interactions (i.e. at small values of $\mathcal{R}$
or small separations between the interacting particles).
Because of this fact numerical results for small systems describe the essential correlations quite well for systems of any size.
Because $V_{\rm QP-QP}(\mathcal{R})$ is not superharmonic at all values of $\mathcal{R}$, Laughlin correlations are
sometimes forbidden.
This results in the absence of Laughlin correlated QP daughter states and of condensed states at certain values of the electron filling
factor like $\nu = 4/11$ and 4/13.
The observation of condensed states at these and other unexpected filling factors immediately suggests pairing
of the CF QP's similar to the pairing of electrons in higher LL's.
This pairing of CF QP's leads to a completely novel set of incompressible states involving new QP's which act like bound states of the
CF excitations despite the repulsive interaction between CF's.

The main thing that we hope you take away from these lectures is that despite the truly beautiful work by
Laughlin and extensions of it by many outstanding scientists (Halperin, Haldane, Wilczek, Schrieffer,
Kivelson, Read, Girvin, MacDonald, Jain, Fradkin, Morf, Chakraborty, Das Sarma, and many others) 
there are always new ways of looking at problems and deeper insights that can lead to interesting new results.

Since the quantum Hall effect involves electrons moving on a two dimensional (2D) surface in the presence of a perpendicular magnetic field, 
we will begin with a brief review of this textbook problem.

\section{Electrons Confined to a Two Dimensional Surface in a Perpendicular Magnetic Field}
The Hamiltonian describing the motion of a single electron confined to the $x-y$ plane in the
presence of a dc magnetic field $\vec B = B \hat z$ is simply 
$H = (2\mu)^{-1} \left[\vec p + \frac{e}{c}\vec A (\vec r )\right]^2$.
The vector potential $\vec A (\vec r )$ in the {\sl symmetric} gauge is given by 
$\vec A (\vec r ) = \frac{1}{2}B (-y \hat x + x \hat y )$.
We use $\hat x$, $\hat y$, and $\hat z$ as unit vectors along the cartesian axes.
The Schr{\"o}dinger equation $(H - E)\Psi (\vec r ) = 0$ has eigenstates\cite{qm}
\be
\Psi_{nm} (r,\phi) = e^{im\phi} u_{nm} (r),
\label{eigenvector}
\ee
\be
E_{nm} = \frac{1}{2} \hbar\omega_c (2n +1 + m +|m|).
\label{eigenvalue}
\ee
\ni
The radial function $u(r)$ satisfies the differential equation
\be
\frac{d^2u}{dx^2} + x^{-1} \frac{du}{dx} - (m^2x^{-1} + x^2 -\epsilon)u = 0
\label{radialeq}
\ee
where $x^2 = (eB/2\hbar c)r^2$ and $\epsilon =(4E/\hbar \omega_c) - 2m$.
The radial wavefunctions can be expressed in terms of associated Laguerre polynomials as
\be
u_{nm} (x) = x^{|m|} \exp(-x^2/2)L_n^{|m|}(x^2).
\label{u_nm}
\ee
\ni
Here $L_0^{|m|} (x^2)$ is independent of $x$ and $L_1^{|m|}(x^2) \propto (|m|+1-x^2)$.
From Eq.(\ref{eigenvalue}) it is apparent that the spectrum of single particle energies consists of highly
degenerate levels with energy $E$ taking on the values $\frac{1}{2}\hbar \omega_c$, $\frac{3}{2}\hbar \omega_c$,
$\cdot\cdot\cdot$.  
These levels are called Landau levels; the lowest LL has $n=0$ and $m=0, -1, -2, \cdot\cdot\cdot$, and
its wavefunction can be written $\Psi_{0m} = z^{|m|}\exp(-|z|^2/4\lambda^2)$, where $z$ stands for $r e^{-i\phi}$
and $ \lambda^2 = \hbar c/eB$.
For a finite size sample of area $\mathcal{A} = \pi R^2$, the number of single particle states in the lowest LL is given by
$N_\phi = B\mathcal{A}/\phi_0$, where $\phi_0 = hc/e$ is the quantum of flux.
The filling factor $\nu$ is defined as $N/N_\phi$, so that $\nu^{-1}$ is simply equal to the number of flux quanta of the 
dc magnetic field per electron.

\section{Integer Quantum Hall Effect}
When $\nu$ is equal to an integer, there is an energy gap (equal to $\hbar \omega_c$) between the filled states and the empty states.
This makes the non-interacting electron system incompressible, because an infinitesimal decrease in the area
$\mathcal{A}$ can be accomplished only at the expense of promoting an electron across the energy gap and into the first unoccupied LL.
This incompressibility is responsible for the integral quantum Hall effect\cite{iqhe}.
To understand the minima in the diagonal resistivity $\rho_{xx}$ and the plateaus in the Hall resistivity $\rho_{xy}$, it is necessary
to understand that each LL, broadened by collisions with defects and phonons, must contain both extended states and localized states\cite{collision}.
The extended states lie in the central portion of the broadened LL, and the localized states in the wings.
As the chemical potential $\zeta$ sweeps through the LL (by varying either $B$ or the particle number $N$), zeros of $\rho_{xx}$ (at $T=0$) and
flat plateaus of $\rho_{xy}$ occur when $\zeta$ lies within the localized states.

A many particle wavefunction at filling factor $\nu=1$ can be constructed by antisymmetrizing the product function
which places one electron in each of the $N$ states with $0\leq |m| \leq N_{\phi}-1$.
It is straightforward to demonstrate that the many particle wavefunction is, for $\nu =1$ 
\be
\Psi_1 (1, 2, \cdot\cdot\cdot, N) = \prod_{<i,j>} z_{ij} ~\exp[-\sum_l |z_l|^2/4\lambda^2 ],
\label{N-1 particle wavefunction}
\ee
where the product is over all pairs $<i,j>$.

\section{Fractional Quantum Hall Effect}
When the filling factor $\nu$ is smaller than unity, the standard approach of placing $N$ particles in the lowest energy single particle states is not
applicable, because more degenerate states than the number of particles are present in the lowest LL.
Laughlin\cite{laughlin} used remarkable physical insight to propose the wavefunction
\be
\Psi_{1/n} (1, 2, \cdot\cdot\cdot, N) = \prod_{<i,j>} z_{ij}^n ~\exp[-\sum_l |z_l|^2/4\lambda^2 ],
\label{N-1/n particle wavefunction}
\ee
for filling factor $\nu = 1/n$,  where $n$ is an odd integer.
The Laughlin wavefunction has the properties that i) it is antisymmetric under interchange of any pair of particles as long as $n$ is odd,
ii) particles stay farther apart and have lower Coulomb repulsion for $n>1$, and iii) because the wavefunction contains terms with $z_i^m$ for
$0 \leq m \leq n(N-1)$, $N_\phi -1$, the largest value of $m$ in the LL, is equal to $n(N-1)$ giving $\nu = N/N_{\phi} \longrightarrow 1/n$ for
large systems in agreement with experiment\cite{fqhe}.
Laughlin also proposed the form of the QP excitations, and evaluated the ground state energy and the gap for creation of
a quasielectron(QE)--quasihole(QH) pair.

\section{Numerical Studies}
Remarkable confirmation of Laughlin's hypothesis was obtained by exact diagonalization\cite{numerical} of the interaction Hamiltonian within
the Hilbert subspace of the lowest LL.
Although real experiments are performed on a 2D plane, it is more convenient to use a spherical 2D surface for numerical diagonalization studies.
The $N$ electrons are confined to a spherical surface of radius $R$.
At the center of the sphere is a magnetic monopole of strength $2Q\phi_0$, where $2Q$ is an integer.
The radial magnetic field is $\vec B = (4\pi R^2)^{-1} 2Q\phi_0 \hat r$.
The single particle states are called monopole harmonics and denoted by $|Q,l,m>$, though we will frequently omit the label $Q$.
The states $|Q,l,m>$ are eigenfunctions of $l^2$ and $l_z$, the square of the angular momentum and its $z$-component
with eigenvalues $l(l+1)$ and $m$ respectively,
as well as of $H_0$, the single particle Hamiltonian, with energy $(\hbar \omega_c/2Q)[l(l+1)-Q^2]$.
Because this energy must be positive, the allowed values of $l$ are given by $l_n = Q + n$, where $n=$ 0, 1, 2, $\cdot\cdot\cdot$.
The lowest LL (or angular momentum shell) has $l_0 = Q$, and a many particle wavefunction can be written as
\be
|m_1, m_2, \cdot\cdot\cdot, m_N > = C_{m_N}^\dagger \cdot\cdot\cdot C_{m_2}^\dagger C_{m_1}^\dagger |0>.
\ee
Here $|m_i| \leq Q$ and $C_{m_i}^\dagger$ creates an electron in state $|l_0,m_i >$.
Clearly there are $G_{NQ} = \left( \begin{array}{c} 2Q+1 \\ N \end{array} \right)$ ways to choose $N$ distinct values of $m$ from the $2Q+1$ allowed values,
so there are $G_{NQ}$ $N$-electron states in the Hilbert space of the lowest LL.
The numerical problem is to diagonalize the interaction Hamiltonian
$H_{INT} = \sum_{i<j} V(|\vec r_i - \vec r_j |)$ in this $G_{NQ}$ dimensional space.
The problem is facilitated by first determining the eigenfunctions $|LM\alpha>$ of the total angular momentum.
Here, $\hat L = \sum_i \hat l_i$, $M=\sum_i m_i$, and $\alpha$ is an additional label that accounts for distinct multiplets with the same total angular momentum.
Because $H_{INT}$ is a scalar, the Wigner--Eckart theorem
\be
<L'M'\alpha'|H_{INT}|LM\alpha> = \delta_{LL'} \delta_{MM'} <L'\alpha'|H_{INT}|L\alpha>
\label{wigner-eckart theorem}
\ee
tells us that matrix elements of $H_{INT}$ are independent of $M$ and vanish unless $L'=L$.
This reduces the size of the matrix to be diagonalized enormously\cite{wigner}.
For example, for $N=10$ and $Q=27/2$ ($\nu =1/3$ state of ten electrons) $G_{NQ} = 13,123,110$ and there are
246,448 distinct $L$ multiplets with $0\leq L \leq 90$.
However, the largest matrix diagonalized is only 7069 by 7069.

Some exact diagonalization results ($E ~{\rm vs} ~L$) for the ten electron system are shown in Fig. 1.
The Laughlin $L=0$ incompressible ground state occurs at $2Q=3(N-1)$ for the $\nu = 1/3$ state.
States with larger values of $Q$ contain one, two, or three QH's ($2Q=28, 29, 30$), and states with
smaller values of $Q$ contain QE's in the ground states.
Figure 1(f) has an $L=0$ ground state corresponding to $\nu=2/5$.

It is probably worth noting that on a plane the allowed values of $m$, the $z-$component of the 
single particle angular momentum, are 0, 1, $\cdot\cdot\cdot$, $N_\phi - 1$.
$M=\sum_i m_i$ is the total $z-$component of angular momentum (the sum is over occupied states).
It can be divided into $M_{\rm CM} + M_{\rm R}$, the center of mass and relative contributions.
The connection between the planar and spherical geometries is $M=Nl+L_z$, $M_{\rm R} = Nl -L$, and 
$M_{\rm CM} = L+L_z$.
The interactions\cite{connection} depend only on $M_{\rm R}$, so $|M_{\rm R}, M_{\rm CM}>$ acts just like $|L,L_z>$.
The absence of boundary conditions and the complete rotational symmetry make the spherical geometry attractive to theorists.
Many experimentalists prefer using the $|M_{\rm R},M_{\rm CM}>$ states of the planar geometry.

%\clearpage
\begin{figure}
 \resizebox{13.0cm}{14.0cm}{\includegraphics{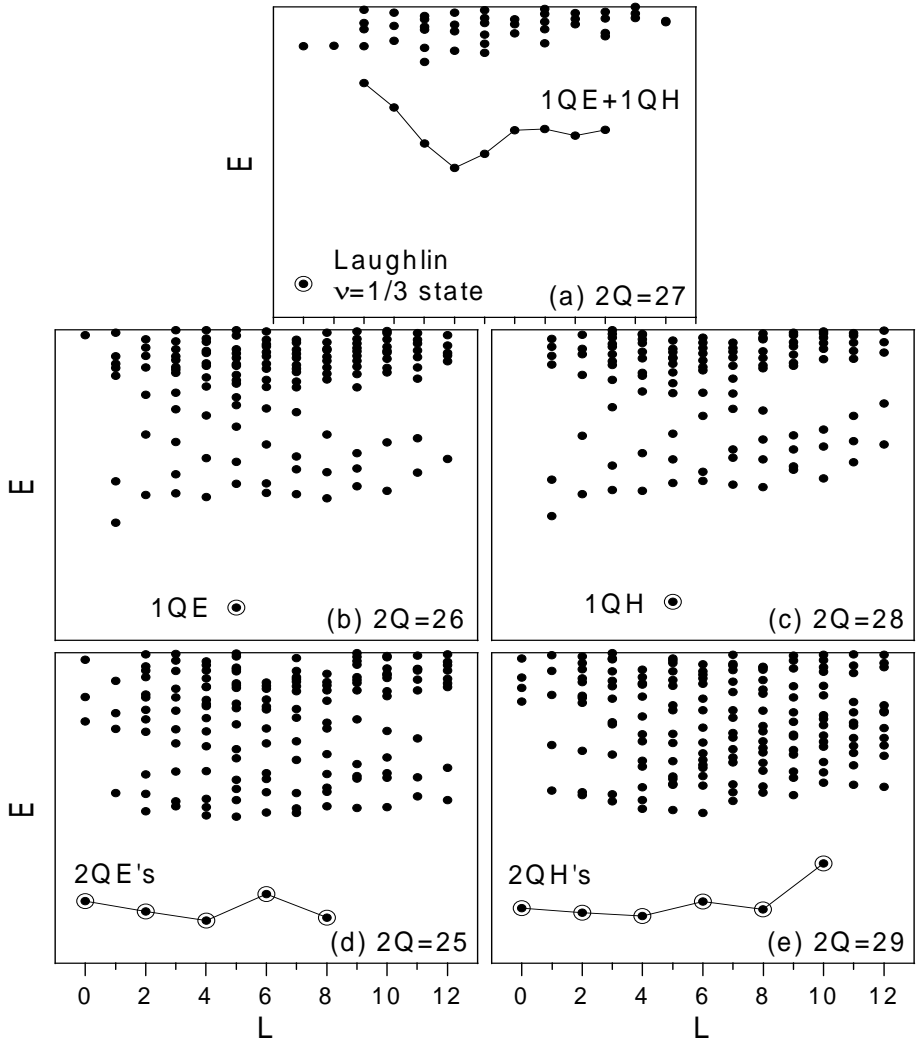}}
\caption{\label{fig1}
   The energy spectra of 10 electrons in the lowest Landau level calculated on a Haldane sphere with
   $2Q$ between 25 and 29.
   The open circles and solid lines mark the lowest energy bands with the fewest composite Fermion 
   quasiparticles.}
\end{figure}

%\clearpage
\section{Chern--Simons Gauge Field}
The Chern--Simons gauge field is introduced by attaching to each electron a flux tube carrying 
a flux $\alpha \phi_0$, where $\phi_0$ is the flux quantum\cite{wilczek}.
This gives rise to a CS magnetic field $\vec b (\vec r )= \alpha \phi_0 \sum_i \delta(\vec r - \vec r_i )\hat z$,
where $\vec r_i$ is the position of the $i^{th}$ electron.
$\vec b (\vec r )$ has no effect on the classical equations of motion because no two electrons occupy the same position, and a given electron
never senses the $\delta$-function magnetic field due to other electrons.
However, in quantum mechanical systems, the vector potential $\vec a (\vec r )$ given by
\be
\vec a (\vec r ) = \alpha \phi_0 \int d^2 r_1 \frac{\hat z \times (\vec r - \vec r_1 )}{|\vec r - \vec r_1 |^2}
\Psi^\dagger (r_1) \Psi(r_1),
\label{gaugefield}
\ee
does introduce a phase factor into the wavefunction\cite{wilczek}.
Here $\Psi^\dagger (r_1)\Psi(r_1)$ is just the density operator $\rho(r_1)$ for the electron fluid.
The Hamiltonian including the CS gauge field $\vec a (\vec r )$ is
\be
H = (2\mu)^{-1} \int d^2 r \Psi^\dagger (r) 
\left[ -i\hbar \vec \nabla_r + \frac{e}{c}\vec A (\vec r ) + \frac{e}{c}\vec a (\vec r ) \right]^2\Psi(r).
\label{total hamiltonian}
\ee
This Hamiltonian contains terms proportional to $\vec a (\vec r )$ to the $n^{th}$ power with $n=0, 1$, or 2.
The $n=1$ term gives rise to a standard two body interaction. 
The $n=2$ term gives three body interactions containing the operator 
$\Psi^\dagger (\vec r )\Psi(\vec r )\Psi^\dagger (\vec r_1 )\Psi(\vec r_1 )\Psi^\dagger (\vec r_2 )\Psi(\vec r_2 )$.
The three body terms are complicated, and they are frequently neglected.

The CS Hamiltonian, usually introduced via a gauge transformation, is considerably more complicated than the original Hamiltonian with
$\vec a (\vec r ) = 0$.
Simplification results only when the MF approximation is made.
This is accomplished by replacing the operator $\rho(\vec r )$ in the CS vector potential [Eq.(\ref{gaugefield})],
by its MF value $n_S$, the uniform equilibrium electron density.
The resulting MF Hamiltonian is a sum of single particle Hamiltonians in which an {\sl effective} magnetic field 
$B^* = B + \alpha \phi_0 n_S$ appears.

\section{Jain's Composite Fermion Picture}
Jain introduced the idea of a composite Fermion (CF) to represent an electron with an attached flux tube which carried 
an even number $\alpha(= 2p)$ of flux quanta\cite{cf}. 
In the MF approximation the CF filling factor $\nu^*$ is given by ${\nu^*}^{-1} = \nu^{-1} - \alpha$, i.e. the number
of flux quanta per electron of the dc field less the CF flux per electron.
When $\nu^*$ is equal to an integer $n=\pm1, \pm2, \cdot\cdot\cdot$, then $\nu = n (1+\alpha n)^{-1}$ generates
(for $\alpha =2$) quantum Hall states at $\nu = 1/3, 2/5, 3/7, \cdot\cdot\cdot$ and $\nu = 1, 2/3, 3/5, \cdot\cdot\cdot$.
These are the most pronounced FQH states observed.

In the spherical geometry one can introduce an effective monopole strength seen by one CF\cite{chen}.
It is given by $2Q^* = 2Q - \alpha (N-1)$ since the $\alpha$ flux quanta attached to every other CF must be subtracted from
the original monopole strength $2Q$.
Then $|Q^*| = l_0^*$ plays the role of the angular momentum of the lowest CF shell just as $Q=l_0$ was the angular
momentum of the lowest electron shell.
When $2Q$ is equal to an odd integer $(1+\alpha)$ times $(N-1)$, the CF shell $l_0^*$ is completely filled, and
an $L=0$ incompressible Laughlin state at filling factor $\nu = (1+\alpha)^{-1}$ results.
When $2|Q^*|+1$ is smaller (larger) than $N$, QE's (QH's) appear in the shell 
$l_{\rm QE} = l_0^* +1$ $(l_{\rm QH} = l_0^* )$.
The low energy sector of the energy spectrum consists of the states with the minimum number of QP excitations required by the value of $2Q^*$ and $N$.
The first excited band of states will contain one additional QE--QH pairs.
The total angular momentum of these states in the lowest energy sector can be predicted by addition of the angular momenta ($l_{\rm QH}$ or $l_{\rm QE}$) of
the $n_{\rm QH}$ or $n_{\rm QE}$ quasiparticles treated as identical Fermions.
In Table I we demonstrated how these allowed $L$ values are found for a ten electron system with $2Q$ in the 
range $29 \geq 2Q \geq 21$. 
By comparing with numerical results presented in Fig. 1, we readily observe that the total angular momentum
multiplets appearing in the lowest energy sector are always correctly predicted by this simple MF CS picture.

It is quite surprising that this MF CS picture works so well.
Fluctuations beyond the MF interact via both Coulomb and CS gauge interactions.
The MF CS picture introduces a new energy scale $\hbar \omega_c^*$ proportional to the effective magnetic field $B^*$, in addition to the Coulomb scale.
For large values of the applied magnetic field, this new energy scale is very large compared with the Coulomb scale, but it is totally irrelevant 
to the determination of the low energy spectrum.
Halperin, Lee, and Read\cite{halperin} treated the interactions beyond the mean field theory for the CF liquid at $B^* = 0$ 
that results from the electron filling factor $\nu = 1/2$.
Lopez and Fradkin\cite{lopez} used the same approach somewhat earlier to treat condensed states at integral CF filling
$\nu^*$.
The many body calculations are usually carried out in the random phase approximation, despite the lack of a small
parameter to justify this approximation.

Jain took a somewhat different approach.
He proposed trial wavefunctions analogous to the Laughlin functions, that are constructed from $\Psi_n$,
the Slater determinants describing the state with $n$ filled LL's, by multiplication by a Jastrow factor
$Z_{2m} = \prod_{j<k} (z_j - z_k)^{2m}$.
This trial function is projected onto the lowest LL at the original magnetic field $B$.
Jain used the trial functions to estimate the energies of the fractional quantum Hall states at $\nu = n(1\pm 2mn)^{-1}$
where $n$ is a positive integer.
Because the trial functions reside within the Hilbert subspace of the lowest LL, 
Jain avoided the superfluous MF energy scale.

\begin{table}
\caption{The effective CF monopole strength $2Q^*$, the number of CF quasiparticles (quasiholes $n_{\rm QH}$ and quasielectrons $n_{\rm QE}$), 
the quasiparticle angular momenta $l_{\rm QE}$ and $l_{\rm QH}$, and 
the angular momenta $L$ of the lowest lying band of multiplets for a ten electron system at $2Q$ between 29 and 21.}
\begin{tabular}{|l|c|c|c|c|c|c|c|c|c|} \hline
$2Q$&29&28&27&26&25&24&23&22&21\\ \hline
%\tableline
$2Q^*$&11&10&9&8&7&6&5&4&3\\  \hline
$n_{\rm QH}$&2&1&0&0&0&0&0&0&0\\  \hline
$n_{\rm QE}$&0&0&0&1&2&3&4&5&6\\   \hline
$l_{\rm QH}$&5.5&5&4.5&4&3.5&3&2.5&2&1.5\\  \hline
$l_{\rm QE}$&6.5&6&5.5&5&4.5&4&3.5&3&2.5\\   \hline
$L$  & 10,8,6,4,2,0 &5 &0 &5 &$8,6,4,2,0$ 
&$9,7,6,5,4,3^2,1$ & $8,6,5,4^2,2^2,0$
&5,3,1 &0 \\ \hline
\end{tabular}
\end{table}

\section{Pseudopotentials}
Electron pair states in the spherical geometry are characterized by a pair angular momentum $L' = L_{12}$.
The Wigner--Eckart theorem tells us that the interaction energy $V_n (L')$ depends only on $L'$ and the LL index $n$.
Figure 2 gives a plot of $V_n (L')$ vs $L'(L'+1)$ for the $n=0$ and $n=1$ LL's.
We define a harmonic pseudopotential $V_H(L')$ to be one that is of the form $V_H = A+BL'(L'+1)$, where $A$ and $B$ are constants.
The allowed values of $L'$ for a pair of Fermions each of angular momentum $l$ are given by $L'=2l-\mathcal{R}$,
where $\mathcal{R}$ is referred to as the {\sl relative} angular momentum and must be an odd integer.
We define $V(L')$ to be {\sl superharmonic} ({\sl subharmonic}) at $L'=2l - \mathcal{R}$ if it increases approaching
this value more quickly (slowly) than the harmonic pseudopotential appropriate at $L' - 2$.
We often write the pseudopotential as $V(\cal{R})$ since $L'=2l-\cal{R}$. 
For the lowest LL $V_0(\cal{R})$ is superharmonic everywhere.
This is apparent for the largest values of $L'$ in Fig. 2.
For the first excited LL $V_1(\mathcal{R})$ is superharmonic only for $\mathcal{R} > 1$.
Although $V_1$ increases between $L'=2l -3$ and $L'=2l-1$, it increases either harmonically or more slowly.
For higher LL's ($n=2, 3, 4, \cdot\cdot\cdot$) $V_n(L')$ increases even more slowly or decreases at
the largest values of $L'$.
The reason for this is that the wavefunctions of the higher LL's have one or more nodes giving structure to the electron charge density.
When the separation between the particles becomes comparable to the scale of the structure, 
the repulsion is weaker than for structureless particles.

%\clearpage
\begin{figure}
\resizebox{13.0cm}{7.0cm}{\includegraphics{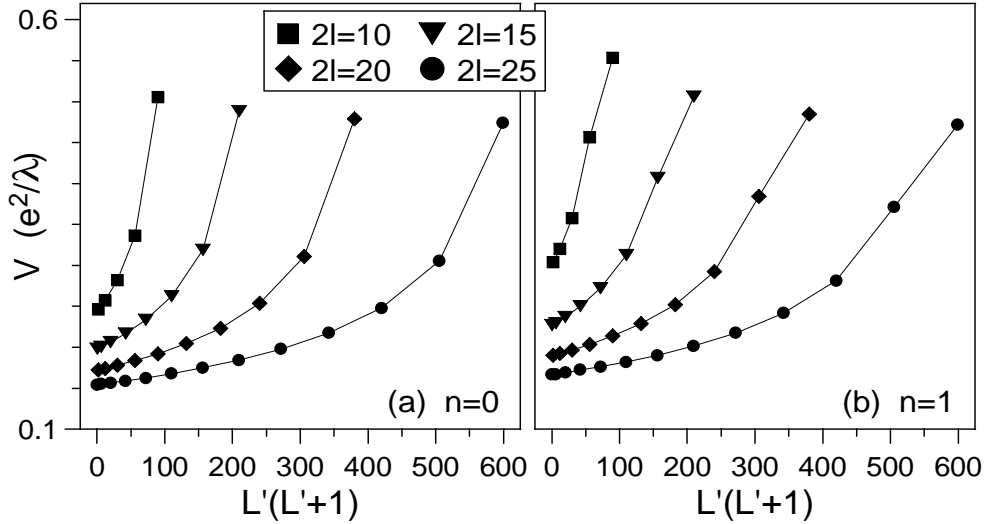}}
\caption{\label{fig2}
   Pseudopotential $V(L')$ of the Coulomb interaction in the lowest (a) and the first excited Landau level (b) as
   a function of squared pair angular momentum $L'(L'+1)$.
   Squares ($l=5$), triangles ($l=15/2$), diamonds ($l=10$), and circles ($l=25/2$) indicate data for different 
   values of $Q=l+n$.} 
\end{figure}

%\clearpage
\section{Angular Momentum}
We have already seen that a spin polarized shell containing $N$ Fermions each with angular momentum $l$ can be described by 
eigenfunctions of the total angular momentum $\hat L = \sum_i \hat l_i$ and its $z$-component $M = \sum_i m_i$.
We define $f_L (N,l)$ as the number of multiplets of total angular momentum $L$ that can be formed from 
$N$ Fermions each with angular momentum $l$.
We usually label these multiplets as $|l^N;L\alpha>$ where it is understood that each multiplet contains $2L+1$
states having $-L \leq M \leq L$, and $\alpha$ is the label that distinguishes different multiplets with the same value of $L$.
We define $\hat L_{ij} = \hat l_i + \hat l_j$, the angular momentum of the pair $i, j$ each member of which has angular momentum
$l$.
The following theorems are quite useful:

\ni
Theorem 1.
\be
{\hat L}^2 + N(N-2){\hat l}^2 = \sum_{<i,j>} {\hat L_{ij}}^2.
\label{theorem1}
\ee
The sum on the right hand side is over all pairs $<i,j>$.

\ni
Theorem 2.
\be
f_L (N,l) \geq f_L (N,l^*),
\label{theorem2}
\ee
where $l^* = l -(N-1)$ and $2l \geq N-1$.

\ni
Theorem 3.

If $g_L (N,l)$ is the Boson equivalent of $f_L (N,l)$, then
\be
g_L (N,l_B) = f_L (N,l_F),
\label{theorem3}
\ee
if $l_B = l_F -\frac{1}{2}(N-1)$.

\ni
The first theorem can be proven very simply using the definitions of ${\hat L}^2$ and $\sum_{<i,j>}{\hat L_{ij}}^2$
and eliminating $\hat l_i \cdot \hat l_j$ from the pair of equations.
The other two theorems are almost obvious conjectures to a physicist, but there exist rigorous 
mathematical proofs\cite{theorem} of their validity.

\section{Coefficients of Fractional Parentage}
Suppose three Fermions each have angular momentum $l$.
One can determine the total angular momentum $L$ by adding $l_1$ and $l_2$ to obtain $L_{12}$ and then adding $l_3$
to obtain $L$.
We can express this as
\be
|l^3;L\alpha> = \sum_{L_{12}} G_{L\alpha}(L_{12})|l^2,L_{12};l;L>
\label{3l}
\ee
$|l^2,L_{12};l;L>$ is a wavefunction in which $l_1$ and $l_2$ have been combined to obtain $L_{12}$.
It is antisymmetric under interchange of the Fermions 1 and 2.
Next one adds the third Fermion with angular momentum $l_3=l$ to obtain $L$.
The function $|l^2,L_{12};l;L>$ is not antisymmetric under interchange of 3 with either 1 or 2.
However, if the coefficient $G_{L\alpha}(L_{12})$ is chosen correctly $|l^3;L\alpha>$ is totally antisymmetric.
$G_{L\alpha}(L_{12})$ is called the coefficient of fractional parentage\cite{shell}, and it is related to the Racah coefficients.

A generalization of the three Fermion problem can be made by writing
\be
|l^N;L\alpha> = \sum_{L'\alpha'} \sum_{L_{12}} G_{L\alpha,L'\alpha'}(L_{12})|l^2,L_{12};l^{N-2},L'\alpha';L>.
\label{Nl}
\ee
$|l^{N-2},L'\alpha'>$ is the $\alpha'$ multiplet of total angular momentum $L'$ of $N-2$ Fermions each with angular momentum
$l$.
From $|l^{N-2},L'\alpha'>$ and $|l^2, L_{12}>$ one can construct an eigenfunction of total angular momentum $L$.
The coefficient $G_{L\alpha,L'\alpha'}(L_{12})$ is called the coefficient of fractional grandparentage\cite{shell}.
It produces a totally antisymmetric eigenfunction $|l^N;L\alpha>$.
Equation (\ref{Nl}) together with the theorem on pair angular momentum, Eq.(\ref{theorem1}), allows us to obtain
the following useful result
\be
L(L+1)+N(N-2)l(l+1) = <l^N;L\alpha|\sum_{<i,j>}{\hat L_{ij}}^2|l^N;L\alpha>.
\label{sumrule1}
\ee
Because Eq.(\ref{Nl}) expresses the totally antisymmetric eigenfunction $|l^N;L\alpha>$ 
as a linear combination of states of well
defined pair angular momentum $\hat L_{ij}$, the right hand side of Eq.(\ref{sumrule1}) can be expressed as
\be
\frac{1}{2}N(N-1)\sum_\alpha \mathcal{G}_{L\alpha} (L_{12}) L_{12}(L_{12}+1).
\label{sumrule2}
\ee
In this expression $\mathcal{G}_{L\alpha} (L_{12})$ is equal to a sum over all $L'\alpha'$ of $|G_{L\alpha,L'\alpha'} (L_{12})|^2$,
and is a measure of the amplitude of pair states with pair angular momentum $L_{12}$ in $|l^N;L\alpha>$.
It is interesting to note that the expectation value of square of the pair angular momentum summed over all pairs is 
totally independent of the multiplet $\alpha$.
It depends only on the total angular momentum $L$.
Because the eigenfunctions $|l^N;L\alpha>$ are orthonormal,
\be
\sum_{L_{12}}\sum_{L'\alpha'} G_{L\alpha,L'\alpha'}(L_{12}) G_{L\beta,L'\alpha'}(L_{12}) = \delta_{\alpha\beta}.
\label{sumrule3}
\ee
From the Eqs.(\ref{sumrule1})-(\ref{sumrule3}) we have two useful sum rules\cite{CFhundrule} 
involving $\mathcal{G}_{L\alpha}(L_{12})$.

They are
\be
\sum_{L_{12}} \mathcal{G}_{L\alpha}(L_{12}) = 1,
\label{sumrule-1}
\ee
and
\be
\frac{1}{2}N(N-1)\sum_{L_{12}} L_{12}(L_{12}+1)\mathcal{G}_{L\alpha}(L_{12}) = L(L+1)+N(N-2)l(l+1).
\label{sumrule-2}
\ee

The energy of the multiplet $|l^N;L\alpha>$ is simply
\be
E_\alpha(L) = \frac{1}{2}N(N-1)\sum_{L_{12}} \mathcal{G}_{L\alpha}(L_{12})V(L_{12}),
\label{energy}
\ee
where $V(L_{12})$ is the pseudopotential appropriate to the interacting particles.
Equation (\ref{energy}) together with our sum rules on $\mathcal{G}_{L\alpha}(L_{12})$ gives the remarkable result that for a harmonic
pseudopotential $V_H(L_{12})$ (as defined in Section 8) the energy $E_{\alpha}(L)$ is totally independent of $\alpha$.
This means that all of the eigenfunctions of the same total angular momentum $L$ have the same energy.
The harmonic pseudopotential introduces no correlations; any linear combination of the eigenstates of the total angular momentum having the 
same eigenvalue $L$ is an eigenstate of the harmonic pseudopotential.

\section{Non-harmonic Pseudopotentials and Correlations}
Because the harmonic pseudopotential introduces no correlations, every multiplet with the same total angular momentum
$L$ has the same energy.
Thus if $V_H = A + BL_{12}(L_{12}+1)$
\be
E_\alpha(L) = N\left[\frac{1}{2}(N-1)A + B(N-2)l(l+1)\right] + BL(L+1),
\label{harmonic energy}
\ee
totally independent of $\alpha$ and increasing with $L$ as $L(L+1)$.
Only the anharmonic part of the pseudopotential $\Delta V(\mathcal{R}) = V(\mathcal{R})-V_H(\mathcal{R})$ lifts
the degeneracy of the multiplets of a given $L$.
If $\Delta V$ is zero except at $\mathcal{R}=1$, then $\Delta V (\mathcal{R}=1)$ is the only energy scale responsible for the correlations.
For positive values of $\Delta V(\mathcal{R}=1)$, it is obvious that the lowest energy states will tend to avoid
pair states with $\mathcal{R}=1$ to the maximum possible extent.
This is exactly what we mean by Laughlin correlations.
If $\Delta V(\mathcal{R}=1)$ is very large, then for the $L=0$ ground state $\mathcal{G}(\mathcal{R}=1)$, the amplitude 
for pairs with $\mathcal{R}=1$, is very small when $2Q \geq 3(N-1)$.
Avoiding $\mathcal{R}=1$ is equivalent to avoiding pair states with $m=1$ in the planar geometry.

Pair states with $\mathcal{R}<2p$ can be avoided by making use of Theorem 2 (Eq.(\ref{theorem2})) to select
subsets of $f_L(N,l)$ by introducing an {\sl effective} Fermion angular momentum $l_p^* = l -p(N-1)$, 
where $p$ is an integer.
For states with $2l=n(N-1)$, where $n= 2p+1$, $l_p^*$ selects a single state with $L=0$
because $2l_p^* +1 = N$.
Then the Laughlin ground state avoids all the pair states with $\mathcal{R} <2p$ to the maximum possible extent.
If $V(\mathcal{R})$ has the property that 
$V(1) \gg V(3) \gg \cdot\cdot\cdot \gg V(2p+1)$,
then the energy spectrum splits into bands as shown in Fig. 3.
In this figure a simple example is given for a system of four electrons.
Start with frame (d) which has $2l=23$, and proceed to (c) through (a)
$2l_1^* = 23-2(N-1) =17$, $2l_2^*=11$, and $2l_3^* = 5$.
For $2l_3^*$, there are only three states ($L=0, 2, 4$) all of which contain two QH's each with angular momentum
$l_{\rm QH}=5/2$.
These states avoid pair states with $\mathcal{R} \leq 1$.
For $2l_2^* = 11$, there are two bands.
The lowest band avoids $\mathcal{R} \leq 1$ while the upper band avoids $\mathcal{R} \leq 3$.
The gap between them (at any given value of $L$) is set by $V(\mathcal{R}=1)$.
For $2l_1^* = 17$, there are three bands that avoid pair with $\mathcal{R} \leq 1$, $\mathcal{R} \leq 3$, and 
$\mathcal{R} \leq 5$.
The starting value of $2l=23$ contains four bands.
The gaps between bands (at a given $L$) depend on $V(\mathcal{R}=1)$ for the two highest bands, $V(\mathcal{R}=3)$ for the second
to third highest, etc.
The average overall dependance of each band on $L(L+1)$ is apparent.
The Hilbert space of the lowest LL splits into
subspaces which avoid pair states with $\mathcal{R} \leq 1$, $\mathcal{R} \leq 3$, $\mathcal{R} \leq 5$,
$\mathcal{R} \leq 7$, etc.
This is typical of Laughlin correlations.

%\clearpage
\begin{figure}
\resizebox{13.0cm}{11.88cm}{\includegraphics{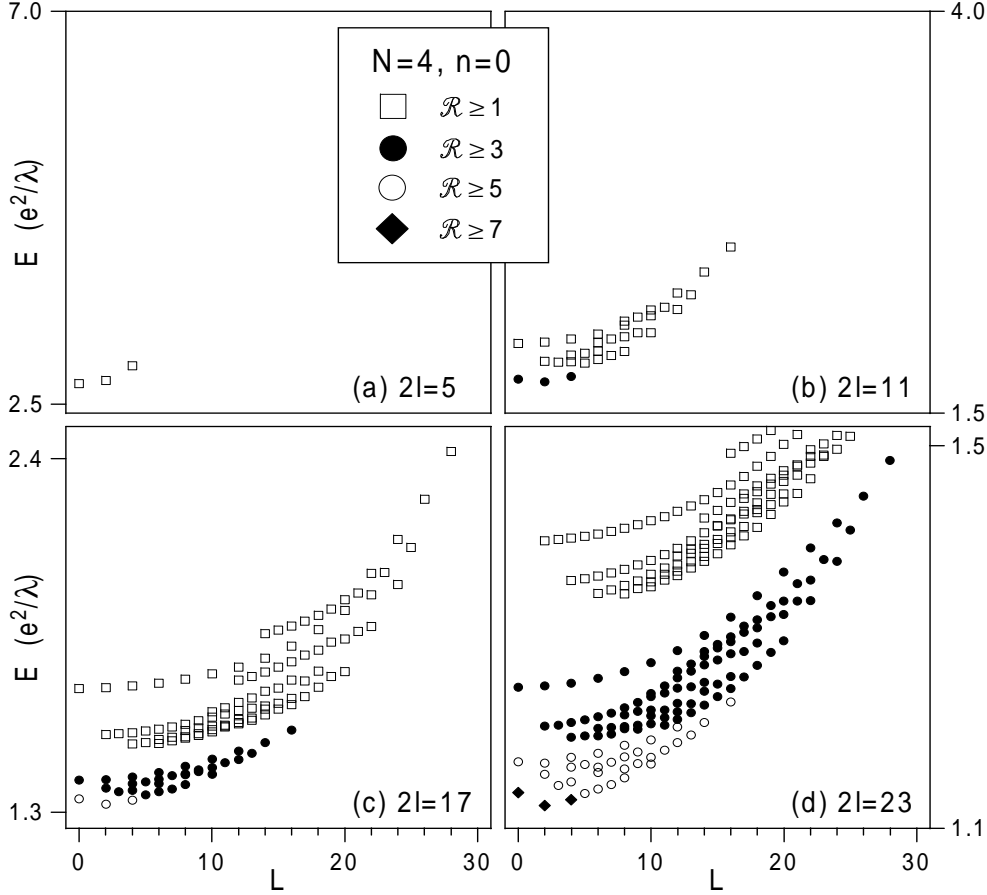}}
\caption{\label{fig3}
   The energy spectra of 4 electrons in the lowest Landau level at different monopole strengths of
   (a) $2Q=5$; (b)$2Q=11$; (c)$2Q=17$; and (d)$2Q=23$. All those $2Q$ values are equivalent in the mean field CF
   picture (CS transformation with $p=$0, 1, 2, and 3 respectively).  
   (solid diamonds: states with $\mathcal{R} \geq 7$, 
   that is $\mathcal{G}(1) \approx \mathcal{G}(3) \approx \mathcal{G}(5) \approx 0$ and $\mathcal{G}(7) >0$;
   open circles: states with $\mathcal{R} \geq 5$, that is $\mathcal{G}(1) \approx \mathcal{G}(3) \approx 0$
   and $\mathcal{G}(5) >0$; 
   solid circles: states with $\mathcal{R} \gg 3$, that is $\mathcal{G}(1) \approx 0$ and $\mathcal{G}(3) >0$;
   open squares: states with $\mathcal{R} \geq 1$, that is $\mathcal{G}(1) >0$).}
\end{figure}

%\clearpage

\section{Correlations in Higher Landau Levels}
For the $n=1$ LL $\Delta V(\mathcal{R}=1)$ is less than or equal to zero.
It is not difficult to see that a state with Laughlin correlations (avoiding $\mathcal{R}=1$ pairs) will have a higher 
energy than one in which some weight $\Delta \mathcal{G} (\mathcal{R}=3)$ is transferred from the value 
of $\mathcal{G}(\mathcal{R}=3)$ for the Laughlin correlated state to $\mathcal{G}(\mathcal{R}=1)$, 
and to states with $\mathcal{R} \geq 5$.
This is caused by the sum rules, Eqs.(\ref{sumrule-1}) and (\ref{sumrule-2}) and the dependence of $E_\alpha(L)$
on $\Delta V (\mathcal{R})$.
We have proposed\cite{pairing} that formation of electron pairs\cite{moore} with $\mathcal{R}=1$ rather 
than Laughlin correlations among
the electrons should occur when $\Delta V(\mathcal{R}) \leq 0$ at $\mathcal{R}=1$.
The pairs can be thought of as Bosons or as Fermions, because in two dimensions a CS transformation interchanges Boson
and Fermion statistics.
It is of more critical importance to realize that if more than a single pair is present, the pair--pair separation
must be sufficiently large that no violation of the Pauli principle is involved when accounting for identical constituent
Fermions belonging to different pairs.
This is accomplished by requiring the largest allowed value of the total angular momentum of two pairs (treated as Fermions) to be
given by $L'=2l_{FP}$, where
\be
2l_{FP} = 2(2l_1 -1) -\gamma_F (N_P-1).
\label{fermionpair}
\ee
Here $l_1$ is the angular momentum of the $n=1$ LL (or shell).
The parameter $\gamma_F$ will be an odd integer (were the pairs treated as Bosons $\gamma_F$ would be replaced by $\gamma_B=\gamma_F -1$), 
and $N_P = \frac{1}{2}N$ is the number of pairs.
The value of $\gamma_F$ is selected  so that the Fermion pair (FP) filling factor
$\nu_{FP} = N_P (2l_{FP} +1)^{-1}$ is equal to unity when the electron filling factor $\nu_1 = N(2l_1+1)^{-1}$
is also equal to unity.
Equation (\ref{fermionpair}) can be thought of as a CS transformation in which $\gamma_F = 3$ 
flux quanta are attached to each Boson pair of angular momentum $2l_1-1$.
We can think of $l_{FP} = 2l_1 -1 -\frac{3}{2}(N_P-1)$ as the {\sl effective} (or mean field) angular momentum of one Fermion pair.
The relation between $\nu_{FP}$ and $\nu_1$ is 
\be
\nu_{FP}^{-1} = 4 \nu_1^{-1} -3
\label{nu_fermionpair}
\ee
for large systems (where terms of the order of $N^{-1}$ can be neglected).

We expect pair formation when the electron filling factor $\nu_1$ of the first excited LL 
satisfies $2/3 \geq \nu_1 \geq 1/3$, 
corresponding to the region in which Laughlin--Jain states\cite{collision,cf}, that avoid $\mathcal{R} =1$ would 
normally occur for a superharmonic potential.
If we assume that all the electrons form pairs, and that
Laughlin correlations occur between different pairs, then incompressible ground states of the FP's
would be expected at $\nu_{FP} =$ 1/3, 1/5, 1/7, and 1/9.
From Eq.(\ref{nu_fermionpair}), these correspond to electron filling factor $\nu_1 =$ 2/3, 1/2, 2/5, and 1/3
respectively.
Only at these values of $\nu_{FP}$ given do we obtain values of $\nu_1$ in the required range.
It is worth noting that $\nu_1 = 2/3$ can be considered one third filled with holes because of electron--hole symmetry.
The FP filling factor for pairs of Fermion holes would be $\nu_{FP}=1/9$ just as it was for the electrons at $\nu_1 =1/3$.
Thus, we can consider the FP filling factors to be small $1/9 \leq \nu_{FP} \leq 1/5$.
At such filling factors, the pair--pair pseudopotential should be Coulomb-like (and superharmonic) since the pair--pair separations are large
compared to the size of a pair.

These ideas can be tested numerically\cite{pairing} by using a model pseudopotential $U_x(\mathcal{R})$ given by $U_x(\mathcal{R}\geq5) = 0$, 
$U_x(\mathcal{R}=1) = 1$, and $U_x(\mathcal{R}=3) = x V_H(\mathcal{R}=3)$ where $V_H(3)$ is the {\sl harmonic} value such that
$U_1$ is linear in $L(L+1)$ between $\mathcal{R}=1$ and $\mathcal{R}=5$.
$U_0$ is superharmonic at $\mathcal{R}=1$ just as the pseudopotential for the lowest LL.
$U_1$ is close to the behavior of the first excited LL, and for $x>2$, $U_x$ is strongly subharmonic.
Remember that we can regard $U_x$ as the anharmonic part of the pseudopotential, to be added to an overall $V_H(\mathcal{R})$.
$V_H$ shifts the energies by $C+BL(L+1)$ (where $C$ and $B$ are constants as given in Eq.(\ref{harmonic energy}))
but doesn't introduce any correlations among multiplets with the same $L$.

In Fig. 4 we present energy spectra for $N=8$ at $2l_1=17$, and $N=10$ at $2l_1=21$ and $2l_1=23$.
Frames (a), (b), and (c) are for the Coulomb interaction in the $n=1$ LL and (d), (e), and (f) are for the model potential
$U_1$.
The $L=0$ ground states correspond to filling factors $\nu_1$ in the first excited LL of 1/3 or 1/2 [(a) and (d)], 
1/2 [(b) and (e)],
and 1/3 [(c) and (f)].
Of course, you must add 2 for the spin up and spin down $n=0$ LL's, which are occupied.
This gives $\nu =$ 5/2 or 7/3, 5/2, and 7/3 for the three cases.
The actual $V_1$ pseudopotential and the model $U_1$ pseudopotential have $L=0$ ground states, 
but the excitations are clearly different.
A term proportional to $L(L+1)$ should be added to the energies in (d), (e), and (f) to account 
for the harmonic contribution to $U_1$.

%\clearpage
\begin{figure}
\resizebox{13.0cm}{9.0cm}{\includegraphics{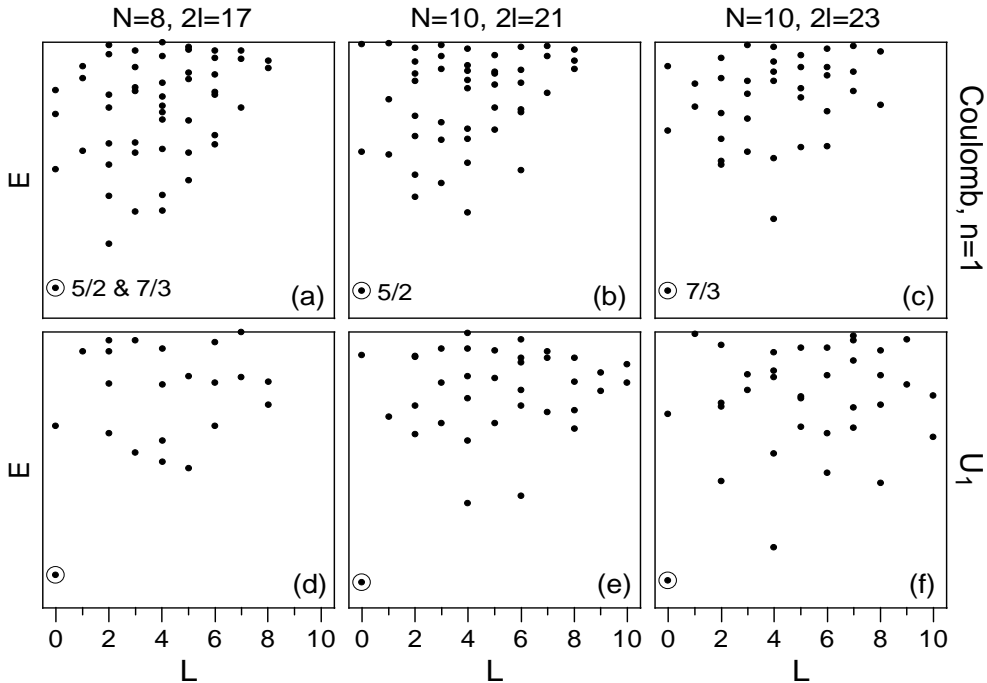}}
\caption{\label{fig4}
   The $N$-electron energy spectra calculated on a Haldane sphere with different values of $2l$ for Coulomb 
   interaction in the $n=1$ Landau level [(a)-(c)] and for model pseudopotential $U_1$ [(d)-(f)].}
\end{figure}

%\clearpage
In Fig. 5 we display $\mathcal{G}(\mathcal{R})$, amplitude for pair states of relative angular momentum $\mathcal{R}$
for the $L=0$ ground states shown in Fig. 4.
For the sake of comparison $\mathcal{G}(\mathcal{R})$ vs $\mathcal{R}$ is shown for the lowest energy states of the 
Coulomb pseudopotential
in the lowest LL for $2l_0$ = 17, 21, and 23.
Notice that $\mathcal{G}(\mathcal{R}=1)$ increases in going from the lowest to the first excited LL,
while $\mathcal{G}(\mathcal{R}=3)$ undergoes a substantial decrease.
{\sl This is evidence of the formation of pairs and the avoidance of the pair state with $\mathcal{R}=3$}.

\begin{figure}
\resizebox{13.0cm}{9.0cm}{\includegraphics{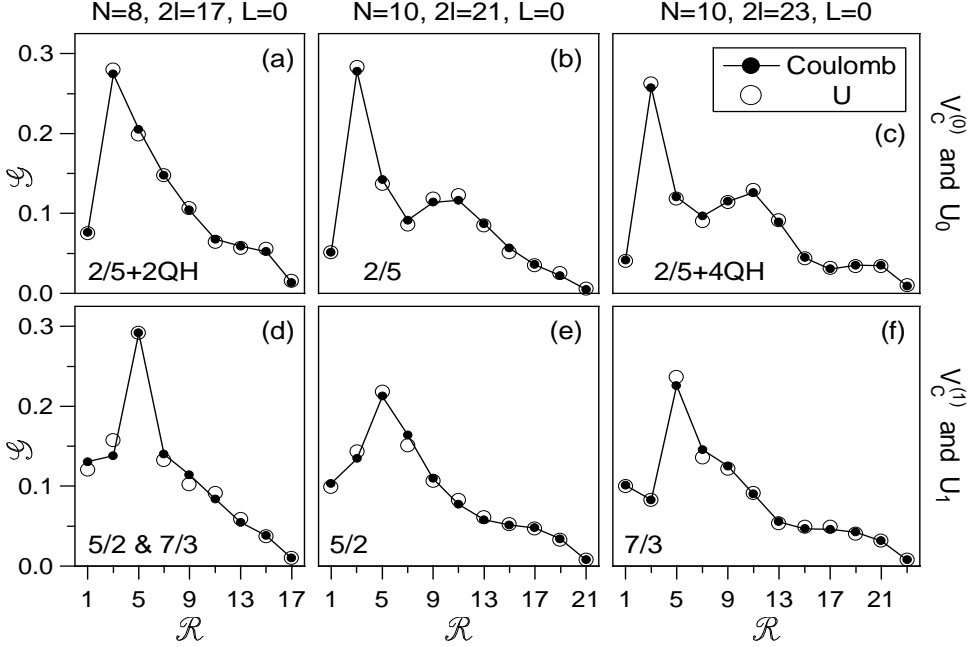}}
\caption{\label{fig5}
   Pair amplitude profiles for the lowest-energy $N$-electron states at $L=0$,
   calculated on a Haldane sphere with different values of $2l$:
   Coulomb interactions in the $n=0$ Landau level (a-c) and in the $n=1$ Landau level (d-f).}
\end{figure}

%\clearpage
It should be noted that the Fermion pair state at $\nu_{FP} = 1/5$ occurs at $2l_{FP} = 5(N_P -1)$
while that at $\nu_{FP} = 1/9$ occurs at $2l_{FP} = 9(N_P-1)$.
Using Eq.(\ref{fermionpair}) for $2l_{FP}$ and setting $\gamma_F=3$ gives the relation\cite{pairing}
between $2l_1$ and $N$ 
for each of the appropriate filling factors $\nu_1$.
For the $\nu_1 = 1/2$ state we find $2l_1 = 2N-3$.
In this case there is a complementary state obtained 
by replacing $N$ by $2l_1+1-N$, i.e. making use of electron--hole symmetry.
It gives for $\nu_1 = 1/2$ the complementary value of $2l_1 =2N+1$.
For the $\nu_1 = 1/3$ state we obtain $2l_1 = 3N-5$.
Even though we find an $L=0$ ground state and an energy gap for excitations at this value,
an $L=0$ ground state with a somewhat larger gap is found at $2l_1=3N-7$.
In Figs. 4 and 5 this value was used and described a $\nu_1 =1/3$ state.
Of course in large systems these finite size corrections are negligible.
A model which considers the $N$ electron system near $\nu_1 = 1/3$ divided into $N_1$ unpaired electrons and
$N_2 = \frac{1}{2}(N-N_1)$ pairs has been studied for some small systems ($N \leq 12 $) with some, but not complete
success.

\section{Chern--Simons Gauge Field Revisited}
In Sections 6 and 7 we introduced the CS gauge field resulting from attaching 
$\alpha$ flux quanta to each electron.
The CS gauge field (or vector potential $\vec a (r)$) was given by Eq.(\ref{gaugefield}), 
and led to a quite complicated Hamiltonian given by Eq.(\ref{total hamiltonian}).
Simplification arose only when the {\sl mean field} approximation was made giving rise to an effective magnetic field
$B^*=B+\alpha \phi_0 n_S$.
The MF CS approximation correctly predicted the structure of the low energy spectrum for any value 
of the applied magnetic field $B$ in a very simple way that involved only the addition of angular momentum of the
minimum number of QE's or QH's required by the monopole strength $2Q$ and the number of electrons $N$.
However, the energy scale of the spectrum involved an effective cyclotron frequency $\omega_c^* = eB^*/mc$ which, 
at large values of $B$, was large compared to the Coulomb interaction, but totally irrelevant to the determination 
of the energy spectrum.
Lopez and Fradkin\cite{lopez} and Halperin, Lee, and Reed\cite{halperin} treated interaction (both Coulomb and CS gauge
interaction) among the fluctuations beyond the mean field by standard many body perturbation theory with
results that were qualitatively correct.
This is somewhat surprising because there is no small parameter to justify simple many body approximations
like the RPA, or even to assure convergence of the perturbation expansion.

We stated earlier that the CS field $\vec b (\vec r ) = \vec \nabla \times \vec a (\vec r )$ had no
effect on the classical equations of motion because the charge on one electron never experience the $\delta$-function
CS magnetic field carried around by the other electrons.
Ever since the classic paper by Aharanov and Bohm\cite{ABeffect} it has become customary to think of quantum mechanical
problems involving a vector potential $\vec a (\vec r )$ within a region in which $\vec b (\vec r )$ vanishes,
in terms of a gauge transformation which alters the phase of the wavefunction but not its magnitude.
A simple example is given in the introductory quantum mechanics text by Griffiths\cite{griffiths}.
Consider a particle of charge $-e$ and mass $m_e$ confined to move on a circular path in the $x-y$ plane.
At the center of the path is a long thin solenoid oriented in the $z$-direction.
When the solenoid carries a current which produces a flux $\Phi = \alpha \phi_0$,
the eigenfunctions $\phi_m (\vec r )$ can be written as
\be
\phi_m (\vec r ) = e^{- (ie/\hbar c) \int \vec a (\vec r )\cdot d\vec r }
\Psi_m (\vec r ),   
\label{CSeigenfunction}
\ee
where $\Psi_m (\vec r ) = e^{im\phi} u_m (r)$ is the eigenfunction given by Eq.(\ref{eigenvector}) when $n=0$.
Because $\vec a (\vec r ) = \left( \frac {\Phi}{2\pi r} \right) \hat \phi $ and
$d\vec r = (\hat r dr + \hat \phi r d\phi)$, the phase factor in Eq.(\ref{CSeigenfunction}) 
can easily be seen to equal $-i \alpha \phi $, where $\phi$ is the angular position of 
the particle along its path.
This means 
\be
\phi_m (\vec r ) = e^{i(m - \alpha )\phi } u_m (r).
\label{transmutation}
\ee
For the case where $\vec r$ describes $\vec r_1 - \vec r_2 $, the relative coordinate of a pair of Fermions,
Eq.(\ref{transmutation}) illustrates the famous transmutation of statistics in 2D systems\cite{wilczek}.
Exchange of particles 1 and 2 corresponds to replacing the angle $\phi$ by $\phi + \pi$.
Therefore, when $\alpha$ is an even integer there is no change in statistics; 
when $\alpha$ is an odd integer Fermions become
Bosons and vice versa, while when $\alpha$ is not an integer the particles obey anyon statistics.
The reason for this is clear.
The extra CS flux enclosed by the unperturbed orbit $u_m$ alters the phase appearing in the angular part of the
wavefunction.
As mentioned in Section 6, the full Hamiltonian including the CS gauge field is, as a first approximation,
usually treated in terms of the mean field.
The qualitative success of this approximation in predicting the structure of the low energy spectrum led
Jain\cite{cf} to suggest a substantial cancellation between the Coulomb and CS gauge interactions beyond the mean field
approximation.
This cannot be true at arbitrary values of the magnetic field $B$, because the two interactions involve
different energy scales with different dependence on $B$.

For a Laughlin $\nu=1/n$ state, where $n$ is an odd integer, the MF CF picture replaces $B$ by $B^*=B/n$.
This changes the magnetic length to $\lambda^* = n^{1/2} \lambda$, and requires the wavefunction given by 
Eq.(\ref{N-1 particle wavefunction}) to contain $\lambda^*$ in place of $\lambda$.
The semiclassical orbit will then have a radius $r_m^* = n^{1/2} r_m = (2n|m|)^{1/2}\lambda$.
For an electron pair in the state with $|m|=1$, the addition of two flux quanta per electron (opposite to $B$)
gives $B^* = B/3$, and the CF orbit with $|m|=1$ would correspond to the electron orbit with $|m|=n$.
Note however that in Eq.(\ref{eigenvalue}), $\omega_c$ is replaced by $\omega_c^*$ which is smaller by a factor 
of $\nu=1/n$.
The lower degeneracy of the CF LL's makes it necessary to consider excited CF states involving the 
energy scale $\omega_c^*$ which is proportional to the magnetic field.
If we look at the limit where $e^2/\lambda \ll \hbar \omega_c$, a correct many body perturbation theory would have
to produce an almost degenerate band of multiplets corresponding to the states of the lowest electron LL,
separated by a huge gap from higher bands corresponding to higher levels.
It is very difficult to see how a many body theory without a small parameter for a perturbation expansion can do this.

A gauge transformation is not the only way in which a vector potential $\vec a (\vec r )$, whose curl vanishes, 
can be introduced into a system.
We can start with the system in some initial state, and slowly increase the value of 
$\alpha \phi_0$, the CS flux,
from zero to the final value where $\alpha$ is some even integer\cite{jj}.
Consider first a pair of electrons with coordinates $\vec r_1$ and $\vec r_2$.
In the single particle Hamiltonian
\be
H_1 = (2\mu)^{-1} \left\{ \vec p_1 + \frac{e}{c}[ \vec A (\vec r_1 ) + \vec a (\vec r_1 )] \right\}^2,
\nonumber
\ee
$\vec A$ is simply replaced by $\vec A + \vec a$.
Here $\vec a (\vec r_1 )= \frac{\alpha \phi_0}{2\pi r} \hat \phi$,
where $r = |\vec r_1 - \vec r_2 |$ and $\hat \phi$ is a unit vector in the direction of increasing relative
coordinate $\phi$.
Note that $\vec a (\vec r_2 ) = - \frac{\alpha \phi_0}{2\pi r} \hat \phi$.
Separating the two particle Hamiltonian into center of mass (CM) and relative (R) coordinate contributions
gives for $H_{\rm R}$
\be
H_{\rm R} = \frac{p^2}{2\mu} + \frac{q\tilde B}{2\mu c} l_z + \frac{q^2 {\tilde B}^2}{8\mu c^2} r^2,
\label{Hrelative}
\ee
where $\tilde B = B + (2\alpha \phi_0/\pi r^2)$.
This is exactly the same equation as that obtained in the absence of CS flux, 
except that the constant $B$ has been replaced by the operator $\tilde B$.
We can assume that the eigenfunction will still be of the form
$\phi_m (r) = e^{im\phi} w_m (r)$, where $w_m(r)$ is a new radial function which must account for the $r$-dependent
terms added to Eq.(\ref{Hrelative}) when $\alpha \neq 0$.
From the term in Eq.(\ref{Hrelative}) proportional to $B$ we obtain a new term 
$\frac{\hbar q m}{2\mu c}\left(\frac{2\alpha \phi_0}{\pi r^2}\right)$;
this can be combined with the $-\frac{\hbar^2}{2\mu}\frac{m^2}{r^2}$ term.
From the term proportional to $B^2 r^2$ in Eq.(\ref{Hrelative}) we obtain two new terms 
$(q^2/2\mu c^2)(\alpha \phi_0/\pi r)^2$ and $(q^2/2\mu c^2 )\alpha \phi_0 B$.
The first of these can also be combined with the $-\frac{\hbar^2}{2\mu}\frac{m^2}{r^2}$ term,
and the second can be combined with the energy $E$.
The result of replacing $B$ by $\tilde B = B + (2\Phi/\pi r^2)$ in Eq.(\ref{Hrelative}) is exactly the same
result one obtains by replacing $m$ by $\tilde m = m+\alpha$.
This means that for $\alpha \neq 0$, the orbit is changed by the CS flux since $u_m(r) \longrightarrow u_{m+\alpha} (r)$.
In contrast to this result, the gauge transformation method of introducing the CS flux would simply give the phase factor in 
Eq.(\ref{CSeigenfunction}) multiplying the radial function $u_m(r)$ for the original electron orbit.
It is important to note that the orbital (i.e. the radial function) is unchanged in the gauge transformation approach, 
but that it changes when the CS flux is added adiabatically.
The angular part of the wavefunction is changed by the Bohm-Aharanov phase factor in the gauge transformation, but
unchanged by the adiabatic addition of CS flux\cite{jj}.
This implies that {\sl there is no change in statistics when the CS flux is added adiabatically.}
Why is this? 
It appears that the time rate of change of the flux through the orbit, which gives rise to an electric field along the orbit
via Faraday's law, causes the relative coordinate $\vec r$ to increase or decrease in magnitude such that
the flux through the orbit remains constant.
Thus if we start with the electron pair in the state $m=-1$ at $\nu=1/3$, and slowly increase the magnitude of the CS flux
from zero to $\Phi = -2 \phi_0$, the final state has a radial wavefunction $u_{|m+\alpha|}(r) = u_3 (r)$.
This orbit encloses three flux units of the dc magnetic field $B$, and minus two CS flux units or the 
same total flux as the original $m=-1$ electron orbit.

It is important to note that in the adiabatic approximation, the CS pair state has the same energy (in the absence of Coulomb repulsion)
as the original electron pair, since all the states of the lowest LL ($m=$0, -1, -2, $\cdot\cdot\cdot$)
are degenerate.
Only when Coulomb interactions are included, do the states with larger average value of $r$ have lower energy.
This is to be contrasted with the mean field CF picture which introduces the new energy proportional to $\hbar\omega_c^*$,
and excitations with energy proportional to $\omega_c^*$ instead of proportional to $e^2/\lambda$, the 
Coulomb scale.

\section{Gedanken Experiment : Laughlin States and the Jain Sequence}
A useful way of arriving at trial functions with built-in Laughlin correlations makes use of the adiabatic 
addition of flux\cite{jj}.
Consider starting with the $\nu=1$ state at a magnetic field $B_{\nu=1}$.
The antisymmetric product state
$\Psi(1, 2, \cdot\cdot\cdot, N) = A \prod_{j=1}^N \psi_{1-j}(z_j)$, where
$\psi_m(z) = e^{im\phi}u_{0m}(r)$ and $A$ is the antisymmetrizing operator, gives us the result appearing 
in Eq.(\ref{N-1 particle wavefunction}) with $\lambda = \lambda_1 = (\hbar c/eB_1)^{1/2}$.
Now adiabatically increase $B$ from $B_{\nu=1}$ to $B_{\nu=1/3} = 3B_1$ while at the same time adiabatically
adding to each electron two CS flux quanta oriented opposite to the dc magnetic field.
The $\nu=1$ wavefunction undergoes two changes:
i) $\lambda_1$ must be replaced by $\lambda_{1/3} = \sqrt{3} \lambda$, and 
ii) a Laughlin--Jastrow factor $\prod_{<i,j>} z_{ij}^2$ must be introduced to account for the effect on
each pair $<i,j>$ of adiabatically
adding the CS flux.
Of course, this is just the Laughlin wavefunction.
The very same idea can be applied to states of the Jain sequence.
For example, consider the $\nu=2/5$ state.
Start with a spin polarized $N$ electron system filling two LL's.
The wavefunction describing this state is an antisymmetric product of single particle wavefunctions
$\Psi_{nm} (\vec r ) = e^{im\phi} u_{nm}(r)$ for $E_{nm}$ in the lowest and first excited LL's.
The radial functions $u_{nm}$ are given by 
$u_{nm} (r) = \left[\frac{n!}{2\pi \lambda_2^2 2^m (n+|m|)!}\right]^{1/2}(\frac{r}{\lambda_2})^{|m|} L_n^{|m|} \left(\frac{r^2}{2\lambda_2^2}\right) \exp\left(-\frac{r^2}{4\lambda_2^2}\right)$,
where $\lambda_2$ is the magnetic length at the magnetic field $B=B_{\nu=2}$, and $L_n^{|m|}$ is an associated
Laguerre polynomial.
Now adiabatically increase $B$ from $B_2$ to $B_{2/5} = 5B_2$ and simultaneously increase the 
magnitude of the CS flux on each electron from $\alpha=0$ to
$|\alpha|=2$.
The result is that $\lambda_2 \longrightarrow \lambda_{2/5} = \sqrt{5} \lambda_2$ in the single particle wavefunctions, and a Laughlin--Jastrow factor
$\prod_{<i,j>} z_{ij}^2$ is introduced.
This is essentially the trial function proposed by Jain, and it is not difficult to see that it resides almost entirely in the Hilbert space
of the lowest LL at $B=B_{2/5}$.

The interesting aspect of the adiabatic introduction of CS flux is that it automatically introduces Laughlin
correlations without resorting to a mean field approximation.
The energies in the absence of Coulomb interactions are totally unchanged, since the trial functions reside in the lowest LL at the
higher dc magnetic field ($B_{1/3}, B_{2/5}, \cdot\cdot\cdot$).
We know that Laughlin correlations by avoiding pair states with the smallest values of $\mathcal{R}$,
select from $f_L(N,l)$ a subset $f_L(N,l^*=l-N+1)$
with smaller repulsion and lower energy only when Coulomb interactions are included.

\section{The Composite Fermion Hierarchy}
Haldane\cite{haldane} introduced the idea of a hierarchy of condensed states in which the Laughlin QP's 
(of the condensed electron states) 
could form {\sl daughter} states, which in turn could have new QP's and on infinitum.
He treated the QP's as Bosons and simply assumed that they would have Laughlin correlations without knowing
much about their {\sl residual} interactions.
The Haldane hierarchy contained, in principle, all fractional filling factors with odd denominators.
Jain's CF picture\cite{cf, jain} gave a simple intuitive picture of certain odd denominator fractions belonging
to the sequence $\nu = n (1+2pn)^{-1}$
where $p$ was a positive integer and $n = \pm 1, \pm2, \cdot\cdot\cdot$.
As discussed earlier, the Jain states can be viewed as integral quantum Hall states of the composite Fermions.
Not all of the fractions belonging to the Haldane hierarchy appear in the Jain sequence, and the relation of the
two hierarchies of incompressible states isn't clear.
The simplest examples are the $\nu = 4/11$ and $\nu = 4/13$ states which belong to the Haldane hierarchy but cannot be
written in the form $n(1+2pn)^{-1}$ for spin polarized states, where $p$ is a positive integer and $n=\pm 1, \pm 2,
\cdot\cdot\cdot$.

Sitko {\sl et al.}\cite{CFhierarchy} introduced the CF hierarchy scheme in order to understand if partially
filled CF shells could give rise to daughter states that would account for the missing odd denominator fractions.
It was suggested that filled CF shells could be ignored, and that 
{\sl the Chern--Simons transformation could simply be applied to the CF QP's 
in a partially filled CF shell}.
This leads to a hierarchy scheme described by the sequence
\be
\nu_l^{-1} = 2p_l + (n_{l+1} + \nu_{l+1})^{-1}.
\label{CFhierarchy}
\ee
Here $\nu_l$ is the QP filling factor at the $l^{th}$ level of the hierarchy, $2p_l$ (where $p_l$ is an integer) is the
number of flux quanta per QP added in the CS transformation to generate the new QP's of the $(l+1)^{st}$ level,
and $n_{l+1}$ is the number of completely filled levels of the new CF QP's.
Of course,
\be
l_{n+1}^* = l_{QP,n}^* - p_n (N_{QP,n} -1)
\label{effective am}
\ee
represents the effective angular momentum of the lowest shell at level $(n+1)$ in terms of angular momentum
$l_{QP,n}^*$ and QP number $N_{QP,n}$ at the $n^{th}$ level of the hierarchy.
If $\nu_{l+1}$ turns out to vanish (i.e. a filled QP shell occurs at level $(l+1)$ of the hierarchy),
then $\nu_l$ is simply the reciprocal of the integer $2p_l + n_{l+1}^{-1}$.
This approach generates all of the Haldane fractions, and it shows the connection between states of the 
Jain sequence and Haldane's continued fractions.

It must be emphasized however, that the electron system has been assumed to be spin polarized, and that the CF QP excitations 
(i.e. CF's in partially filled angular momentum shells) have been assumed to support Laughlin correlations.
The CF hierarchy picture seems a much more reasonable assumption and certainly predates the idea of directly
adding two flux quanta to some electrons and four flux quanta to others to obtain the so called $\rm CF^2$ and
$\rm CF^4$ composite Fermions\cite{pan,park}.

As an illustration let's look at $N=8$ electrons at $2l_0 = 18$.
The CF transformation gives the effective CF angular momentum at the first hierarchy level as
\be
2l_1^* = 2l_0 - 2p_0 (N-1).
\label{l1 star}
\ee
Taking $p_0 = 1$ gives $2l_1^* = 4$.
This CF shell can accommodate $2l_1^* +1 = 5$ composite Fermions leaving $N_{\rm QE,1} =3$ quasielectrons in the 
next shell with angular momentum $l_{\rm QE,1}^* = l_1^* + 1 = 3$.
Reapplying the CS transformation to these three QE's gives
\be
2l_2^* = 2(l_1^* +1) - 2p_1 (N_{\rm QE,1} -1).
\label{l2 star}
\ee
Taking $p_1 =1$ gives $2l_2^* = 6 - 2(3-1) = 2$.
This shell can exactly have $2l_2^* + 1 = 3$ of the QE's giving $n_2 = 1$ and $\nu_2 = 0$.

\be
\nu_1^{-1} = 2p_1 +(n_2 + \nu_2 )^{-1} = 3,
\label{nu1}
\ee
Then,
the hierarchy equations give
\be
\nu^{-1} = 2p_0 +(n_1 + \nu_1 )^{-1} = 11/4,
\label{nu}
\ee
predicting a Laughlin correlated spin polarized state at $\nu = 4/11$ arising from the QE state at $\nu_{\rm QE} =1/3$.

When numerical calculations were carried out for $N = 8$ and $2l_0 = 18$, the low energy spectrum contained the 
five multiplets $L =$ 0, 2, 3, 4, and 6 resulting from the three QE's each with $l_{\rm QE} = 3$.
However, $L = 0$ and $L=3$ states clearly had the highest energies, and the degenerate multiplet at $L=2$ 
had a very slightly lower energy than the multiplets at $L=4$ and 6.
This is illustrated in Fig. 6, a plot of the low energy spectrum of an eight electron system in the lowest
LL at angular momentum $l_0 = 9$.
It was quickly realized that not all the states predicted by the CF hierarchy approach (or by the Haldane 
hierarchy scheme) were realized, but the reason why was not completely clear.
%\clearpage
\begin{figure}
\hspace{2.5cm}
\resizebox{7.0cm}{7.0cm}{\includegraphics{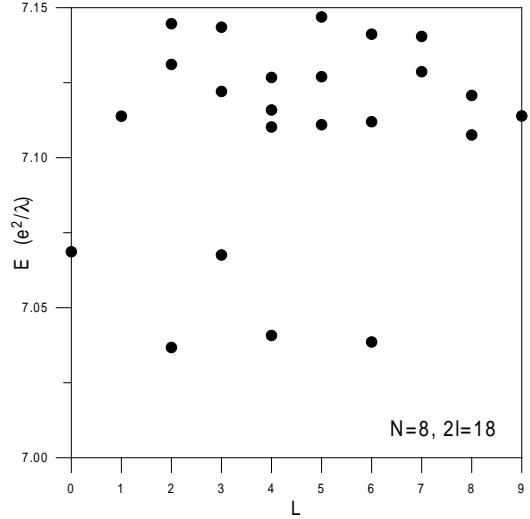}}
\caption{\label{fig6}
   Low energy states of the spectrum of eight electrons at $2l=18$.
   The lowest energy band contains three QE's each with $l_{\rm QE}=3$.
   Reapplying the CS mean field approximation to these QE's would predict an $L=0$ daughter state
   corresponding to $\nu=4/11$.
   The data makes it clear that this is not valid.}
\end{figure}

%\clearpage
\section{Quasiparticle--Quasiparticle Interactions}
Although the mean field CS approximation correctly predicted the structure of the lowest band of states
in the energy spectrum for any value of the dc magnetic field, it was clear that {\sl residual} QP--QP interactions were 
present.
In Fig. 1, frames (d) and (e) show states containing two QE's and two QH's respectively.
In frame (d) each QE has $l_{\rm QE} = 9/2$, so the allowed values of a QE pair angular momentum are 0, 2, 4, 6, and 8.
For the QH's $l_{\rm QH} = 11/2$ and the allowed pair angular momenta are 0, 2, 4, 6, 8, and 10.
If there were no residual QP--QP interactions all the 2QE states would have the same energy, namely $2\varepsilon_{\rm QE}$,
twice the energy of a single QE.
The same would be true for the QH pair, all angular momentum states would have energy $2\varepsilon_{\rm QH}$
in the absence of residual interactions.
Because the pair angular momentum $L'$ is equal to $2l - \mathcal{R}$, 
we can immediately obtain $V_{\rm QE-QE}(\mathcal{R})$
and $V_{\rm QH-QH}(\mathcal{R})$ from the numerical data up to an overall constant 
(which is of no importance in determining the QP--QP correlations).
In Fig. 7 we display $V_{\rm QP-QP}(\mathcal{R})$ as a function of $\mathcal{R}$ obtained from exact numerical diagonalization of 
systems containing up to eleven electrons.
We have considered QP's of the Laughlin $\nu=1/3$ and $\nu=1/5$ states.
Notice that the behavior of QE's is similar for $\nu=1/3$ and $\nu=1/5$ states, and the same is true for QH's of the 1/3 and 1/5
Laughlin states.
Because $V_{\rm QE-QE}(\mathcal{R}=1) < V_{\rm QE-QE}(\mathcal{R}=3)$, and
$V_{\rm QE-QE}(\mathcal{R}=5) < V_{\rm QE-QE}(\mathcal{R}=7)$, we can readily ascertain that $V_{\rm QE-QE}(\mathcal{R})$ is subharmonic at
$\mathcal{R}=1$ and $\mathcal{R}=5$.
Similarly, $V_{\rm QH-QH}(\mathcal{R})$ is subharmonic at $\mathcal{R}=3$ and probably at $\mathcal{R}=7$.

%\clearpage
\begin{figure}
\resizebox{13.0cm}{9.8cm}{\includegraphics{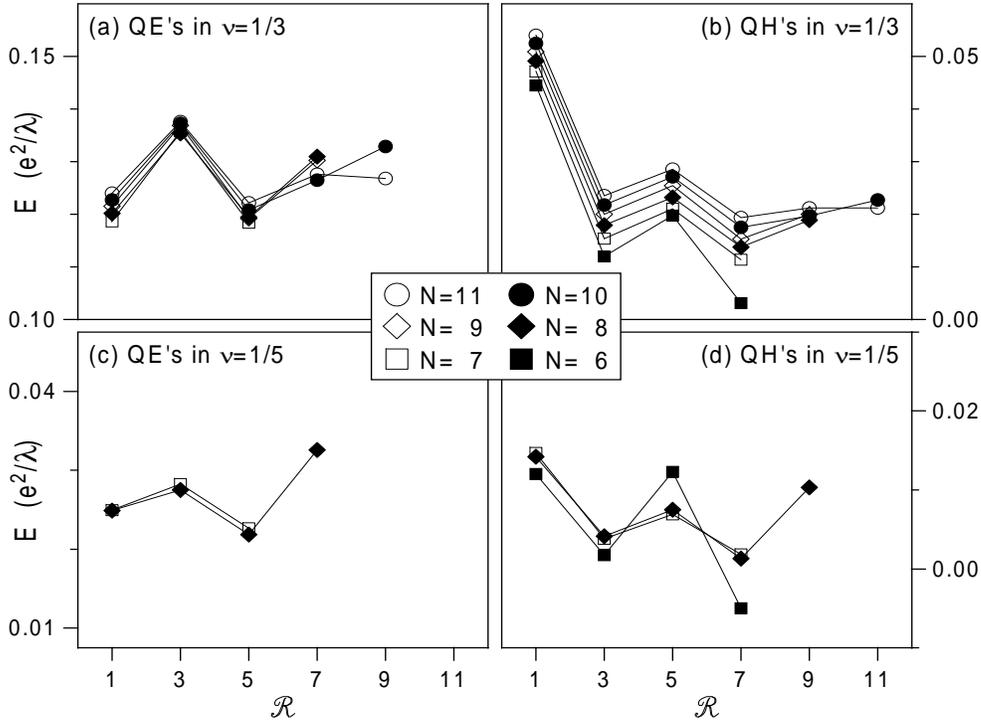}}
\caption{\label{fig7}
   The pseudopotentials of a pair of quasielectrons (left) and quasiholes (right) in Laughlin $\nu=1/3$ (top) and $\nu=1/5$ 
   (bottom) states, as a function of relative pair angular momentum $\mathcal{R}$.
   Different symbols mark data obtained in the diagonalization of between six and eleven electrons.}
\end{figure}

%\clearpage
There are clearly finite size effects since $V_{\rm QP-QP}(\mathcal{R})$ is different for different values of the 
electron number $N$.
However, when plotted as a function of $N^{-1}$, $V_{\rm QP-QP}(\mathcal{R})$ converges rather well to a rather well
defined limit as shown in Fig. 8 for $V_{\rm QE-QE}(\mathcal{R})$ at $\mathcal{R}=$ 1, 3, and 5.
The results are quite accurate up to an overall constant (which is of no significance when you are interested
only in the behavior of $V_{\rm QP-QP}$ as a function of $\mathcal{R}$).
Because the short-range interactions (i.e. at small values of $\mathcal{R}$ or small QP--QP separations) determine
the nature of the ground state, numerical results for small systems describe the important correlations very well for 
systems of any size.

From our discussion of correlations in the first excited LL, it is apparent that Laughlin correlations
among QE's will not occur at $\mathcal{R}=1$ and at $\mathcal{R}=5$,
nor will they occur among QH's at $\mathcal{R}=3$.
This immediately tells us that it is impossible for $\nu_{\rm QE}=1/3$ (and 1/7) and $\nu_{\rm QH}=1/5$ to lead to incompressible
daughter states of the CF hierarchy.
We emphasize that this statement means that 
{\sl for a spin polarized state in which QP's of the Laughlin 
$\nu = 1/3$ state (or $\nu=1/5$ state) yield filling factor $\nu_{\rm QE}=1/3$ (or $\nu_{\rm QH}=1/5$),
Laughlin correlations among the QP's giving rise to daughter states, at e.g. $\nu=4/11$ (or $\nu=4/13$), 
cannot occur!}
How then can we possibly understand the observations\cite{pan} of incompressible states at 
both $\nu=4/11$ and $\nu=4/13$?

\begin{figure}
\hspace{1.5cm}
\resizebox{10.0cm}{7.8cm}{\includegraphics{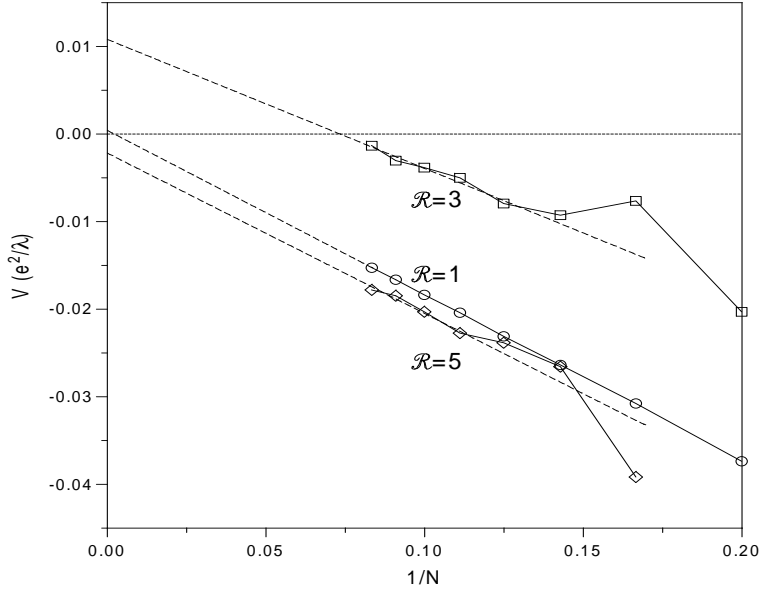}}
\caption{\label{fig8}
   Quasielectron pseudopotential $V_{\rm QE-QE}(\mathcal{R})$ as a function of $N^{-1}$, the inverse of the particle 
   number for the values of relative angular momenta $\mathcal{R} =$ 1, 3, and 5.   
   Extrapolation to $N^{-1} \longrightarrow 0$ corresponds to an 
   infinite planar system.}
\end{figure}

%\clearpage
\section{Quasiparticle--Quasiparticle Pairing and Novel Families of Incompressible States}
If Laughlin correlations among the QP's are ruled out at certain values of the QP filling, 
the observation of incompressible states at such values of $\nu_{\rm QP}$ must be associated 
with one of two possibilities.  
We assumed at the start that the electron system was spin polarized.
It could be that the CF excitations in the partially filled CF shells have their spins reversed 
with respect to the majority spin CF's filling the fully occupied LL's.
Then, if the reversed spin CF's had interactions that gave a superharmonic pseudopotential,
Laughlin correlations among the new reversed spin CF QP's could occur in daughter states.
This was suggested by Park and Jain\cite{park}, as a possible explanation of the 4/11 incompressible state,
but they did not investigate $V_{\rm RQE-RQE}(\mathcal{R})$, the pseudopotential for the interaction of reversed 
spin QE to prove that it could really occur.
Szlufarska {\sl et al.}\cite{reversedspin} did investigate the interaction of reversed spin QE's and demonstrated 
that $V_{\rm RQE-RQE}(\mathcal{R})$ was weakly superharmonic at $\mathcal{R}=1$, making spin unpolarized states a possible
explanation of daughter states that would otherwise be forbidden.

A more interesting possibility is that of pairing\cite{CFpairing} of the CF QP's.
We know from our study of correlations in the first excited LL that pairing of the electrons is expected 
when $V_1(\mathcal{R})$, the pseudopotential describing the interaction of a pair of electrons,
is not superharmonic.
This is exactly the behavior we found for $V_{\rm QP-QP}$ at certain values of $\mathcal{R}$ in Section 16.

Because $V_{\rm QE-QE}(\mathcal{R})$ has its maximum at $\mathcal{R}=3$, the QE's tend to form pairs with $\mathcal{R}=1$ in 
order to minimize the pair amplitude at $\mathcal{R}=3$.
The pairs of {\sl effectively} bound Fermions would usually be treated as Bosons, but in 2D systems Boson and
Fermion statistics can be interchanged via a CS transformation\cite{wilczek,theorem}.
A single pair will have an angular momentum $L'=2l-1$
(the largest possible angular momentum of two Fermions each with angular momentum $l$), 
and a relative angular momentum $\mathcal{R}=1$.
However, when more than a single pair is present, the allowed values of the total angular momentum  of the pair must be 
chosen in such a way that the Pauli principle is not violated when accounting for identical constituent Fermions belonging
to different pairs.
This can be accomplished in exactly the same way as was done for electrons in the first excited LL.
If the pairs are treated as Fermions, the minimum value of the allowed total angular momentum of two pairs is
taken to be $L' = 2l_{FP}$ where
\be
2l_{FP} = 2(2l-1) -\gamma_F (N_P-1).
\label{lFP}
\ee
Here $N_P$ is assumed to equal $N/2$ ($N=$ the number of QE's, each with angular momentum $l$),
and $\gamma_F$ is taken to be 3.
Equation (\ref{lFP}) is exactly what we obtain by a CS transformation that attaches three flux quanta to each pair.
The {\sl effective} (mean field) angular momentum of a single FP is
$l_{FP} = 2l-1 -\frac{3}{2}(N_P-1)$.
The relation between $\nu_{FP}$ and $\nu_{\rm QE}$ is exactly the same as we found in Section 12,
\be
\nu_{FP}^{-1} = 4\nu_{\rm QE}^{-1} -3.
\label{nuFP}
\ee
In Eq.(\ref{nuFP}) terms of order $N^{-1}$ have been omitted since they vanish in the limit of large systems.
The CS transformation given by Eq.(\ref{lFP}) automatically forbids states of two FP's with the smallest separation.
The smallest allowed value of $2l_{FP}$ avoids a violation of the Pauli principle,
and helps the 
individual QE to avoid the largest QE--QE repulsion at $\mathcal{R}=3$.
The CS transformation selects from $g_L (N,l)$, the number of multiplets of total angular momentum $L$ that can be obtained 
from $N$ Fermions each with angular momentum $l$, $g_L(N_P,l_{FP})$ which is a subset\cite{jennifer} of $g_L(N,l)$.

We expect pair formation for QE filling factors satisfying $2/3 \geq \nu_{\rm QE} \geq 1/3$, 
where Laughlin--Jain states that would avoid $\mathcal{R}=1$ occur for a superharmonic pseudopotential.
Fermion pair states with $\nu_{FP} =$ 1/3, 1/5, 1/7, and 1/9 give through Eq.(\ref{nuFP}) QE fillings of 
$\nu_{\rm QE} =$ 2/3, 1/2, 2/5, and 1/3.
Only these values of $\nu_{FP}$ give Laughlin states of the FP's with $\nu_{\rm QE}$ in the required range.
In the hierarchy scheme describing partially filled CF levels, the original electron filling factor is given by
$\nu^{-1} = 2 + (1+\nu_{\rm QE})^{-1}$.
This is just Eq.(\ref{CFhierarchy}) with $p_0=1$, $n_1=1$, and $\nu_1 = \nu_{\rm QE}$.
For QH's, the pairs have $\mathcal{R}=3$ and avoid $\mathcal{R}=5$; we expect them to occur for $\nu_{\rm QH}$
satisfying $1/3 > \nu_{\rm QH} \geq 1/5$.
This novel scheme of QP pairing leads to a novel family of incompressible states as shown in Table II.
\begin{table}
\caption{Novel family of incompressible states resulting from pairing of composite Fermion quasiparticles in the lowest Landau level}
\begin{narrowtabular}{1cm}{l||c|c|c|r} \hline 
$\nu_{FP}$&1/3 &1/5 &1/7 &1/9 \\ \hline \hline
%\tableline
$\nu_{\rm QE}$&2/3 &1/2 &2/5 &1/3 \\  \hline
$\nu$&5/13 &3/8 &[7/19] &4/11 \\  \hline
$\nu_{\rm QH}$&2/7 &1/4 &2/9 &1/5 \\   \hline
$\nu$&5/17 &3/10 &[7/23] &4/13 \\  \hline
\end{narrowtabular}
\end{table}
All of these states except the 7/19 and 7/23 states have been observed.\cite{pan}
We don't know if these states are simply difficult to observe (and might be seen in future experiments) or if there is
some reason why they are not realized in real systems.
We have considered only complete pairing of all the QP's, and this may be an oversimplification that
needs to be reconsidered.

\begin{figure}
\resizebox{13.0cm}{6.28cm}{\includegraphics{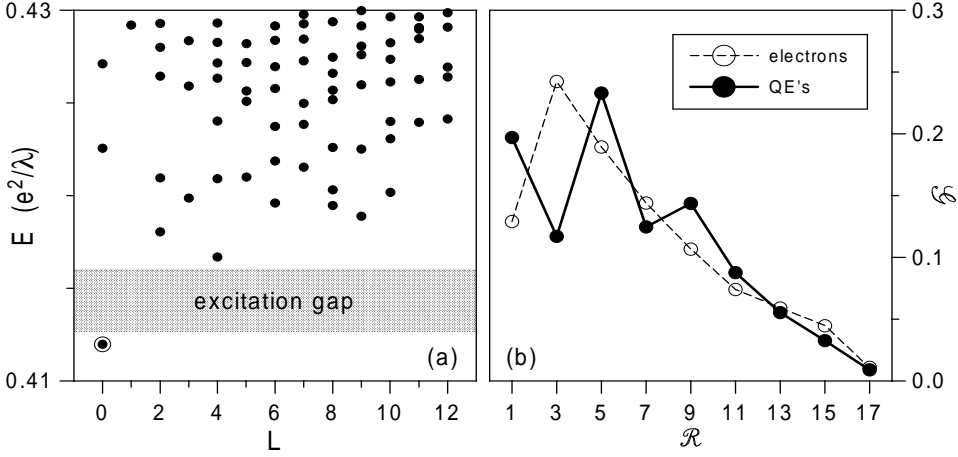}}
\caption{\label{fig9}
   (a)	Energy spectra as a function of total angular momentum L of 10 QE's at $2l=17$ corresponding 
   to $\nu_{\rm QE}=1/2$ and $\nu=3/8$.  
   It is obtained in exact diagonalization in terms of individual QE's interacting 
   through the pseudopotential shown in Fig. 1.
   (b)	Coefficients $\mathcal{G}(\mathcal{R})$, the amplitude associated with pair states of 
   relative angular momentum $\mathcal{R}$, for the lowest $L=0$ state.   
   The solid dots are for 10 QE's of the $\nu=1/3$ state in a shell of angular momentum $l = 17/2$.  
   The open circles are for 10 electrons in the lowest Landau level at $l_0=17/2$.}
\end{figure}

As an illustration we have performed an exact diagonalization on a system containing $N=10$ QE's at
$2l=17$.
This corresponds to $\nu_{\rm QE}=1/2$ and $\nu = 3/8$.
The energy spectrum is given in Fig. 9(a).
It is obtained using the QE--QE pseudopotential presented in Fig. 2.
The $L=0$ ground state is clearly separated by a gap from the lowest excited states.
In Fig. 9(b) we show the behavior of the amplitude $\mathcal{G}(\mathcal{R})$ for the $L=0$ ground state for
all the allowed values of $\mathcal{R}$.
This data was obtained using the QE--QE pseudopotential given in Fig. 7.
For comparison $\mathcal{G}(\mathcal{R})$ is presented for the pseudopotential of electrons in the lowest LL
with $N=10$ and $2l_0=17$.
This state corresponds to a Jain $\nu=3/5$ state containing two QH's each with $l_{\rm QH}=5/2$.
The three states in the low energy sector have $L=0$, 2, and 4, and $\mathcal{G}(\mathcal{R})$ is shown 
for the $L=0$ state.
It should be emphasized that the three low energy states of the superharmonic potential have $\mathcal{G}(\mathcal{R}=3)$
as a maximum and $\mathcal{G}(\mathcal{R}=1)$ as a minimum.
This is typical of Laughlin correlated states for $1/2 \geq \nu \geq 1/3$.
In contrast the subharmonic pseudopotential displays, relative to the superharmonic one, a much larger value of 
$\mathcal{G}(\mathcal{R}=1)$ and a much smaller value of $\mathcal{G}(\mathcal{R}=3)$.
This is in accord with the formation of pairs with $\mathcal{R}=1$ and the avoidance of the maximum QE repulsion
at $\mathcal{R}=3$.
Figure 10 shows the $\mathcal{G}(\mathcal{R})$ values for a more dilute QE state with $N=10$ and $2l=27$.
It is contrasted with the $L=0$ state of the superharmonic potential at $N=10$ and $2l_0=27$.
Note that the $\mathcal{G}(\mathcal{R}=1)$ is roughly equal to 1/9 corresponding to only five $\mathcal{R}=1$ pairs
out of forty five possible pair states.

\begin{figure}
\hspace{1.5cm}
\resizebox{10.0cm}{8.12cm}{\includegraphics{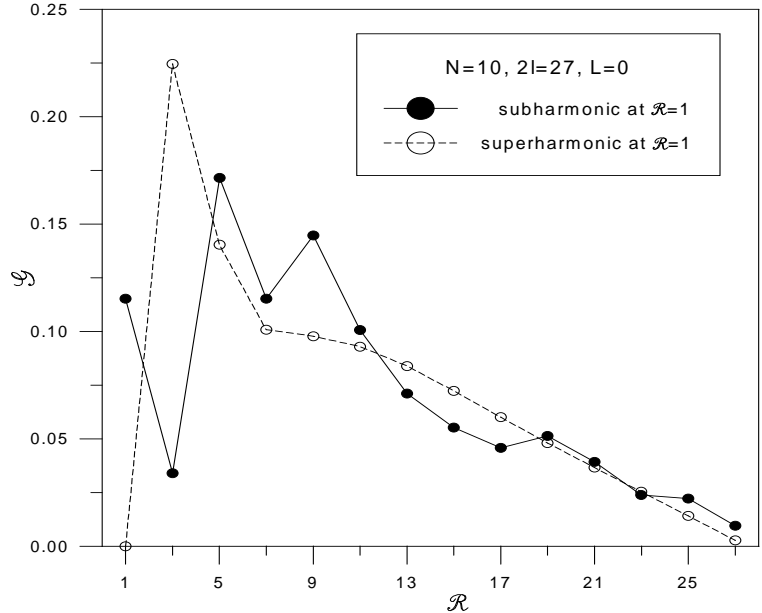}}
\caption{\label{fig10}
   Coefficients $\mathcal{G}(\mathcal{R})$, the amplitude associated with pair states of 
   relative angular momentum $\mathcal{R}$, for the lowest $L=0$ state.   
   The open circles for 10 QE's in a shell of angular momentum $l = 27/2$.  
   The solid dots are for 10 electrons in the lowest Landau level at $l_0=27/2$.}
\end{figure}

%\clearpage

Rather than contradicting the assertion that the 4/11 and 4/13 states cannot be incompressible states 
{\sl of the spin polarized CF hierarchy}, the results of Pan {\sl et al.}\cite{pan} offer support for the idea
of pairing at certain values of $\nu_{\rm QP}$ where Laughlin correlations cannot be supported.
The pairing gives a bonus in that it explains the occurrence of even denominator fractions 
in the lowest LL (at 3/8 and 3/10).
As far as we know, none of the other hierarchy schemes studied so far do this.
We emphasize that the simple repetition of Laughlin correlations among daughter states containing QP's isn't
always appropriate\cite{smet,mani} because of the form of 
$V_{\rm QP-QP} (\mathcal{R})$\cite{numerical,haldane,CFhierarchy,pan,fermiliquid,pseudo,mandel,no411}.
The proposed pairing of CF QP's together with the Laughlin correlations among the pairs gives rise 
to a novel type of QP and to an entirely new hierarchy of incompressible states.

\ni
The authors gratefully acknowledge the support by Grant DE-FG 02-97ER45657 of 
the Material Research Program of Basic Energy Sciences--US Department of Energy.
AW acknowledges support from Grant 2P03B02424 of the Polish KBN 
and KSY acknowledges partial support of the ABRL(R14-2002-029-01002-0) through the KOSEF.

%\section*{Footnotes and References}

\end{document}